\documentclass[aps,prd,noshowpacs,nofootinbib,superscriptaddress,groupedaddress, floatfix,superscriptaddress, preprintnumbers,twocolumn]{revtex4}  
 
\usepackage{graphicx} 
\usepackage{dcolumn}   
\usepackage{bm}        
\usepackage{amssymb}   

\usepackage{hyperref}
\usepackage{multirow}
\usepackage{feynmf}
\usepackage{amsmath,amssymb}

\usepackage{graphicx,bm}

\usepackage{slashed}

\usepackage{natbib}

\usepackage{epstopdf}

\usepackage{ulem} 

\usepackage[usenames]{color}

\usepackage{float}

\renewcommand\sout{\bgroup \color{red} \ULdepth=-.5ex \ULset}

\newcommand{\be}{\begin{equation}}
\newcommand{\ee}{\end{equation}}
\newcommand{\ba}{\begin{eqnarray}}
\newcommand{\ea}{\end{eqnarray}}

\begin{document}
\begin{flushright}
\end{flushright}

\preprint{INT-PUB-21-008}

\title{Quarkyonic Effective Field Theory,  Quark-Nucleon Duality and Ghosts}

\author{Dyana C. Duarte}
\email[]{dyduarte@uw.edu}
\author{Saul Hernandez-Ortiz}
\email[]{saulhdz@uw.edu}
\author{Kie Sang Jeong}
\email[]{ksjeong@uw.edu}
\author{Larry D. McLerran}
\email[]{mclerran@me.com}
\affiliation{Institute for Nuclear Theory, University of Washington, Box 351550, Seattle, WA, 98195, USA}

\date{\today}

\begin{abstract}
We present a  field theoretical description of quarkyonic matter consisting of quark, nucleon and ghost fields coupling to mesonic degrees of freedom.  The ghosts are present to cancel over-counting of nucleon states that are Pauli blocked by the quark Fermi sea.  Such a theory becomes an effective field theory of nucleons at low baryon density, and as such will reproduce nucleonic matter phenomenology.  This theory can accommodate chiral symmetry restoration and the dynamical generation of a shell of nucleons at the Fermi surface.  It is valid for finite temperature and density.  In such a theory, quark-nucleon duality is accomplished by inclusion of ghost fields so that the nucleons extra degrees of freedom, that are beyond those of quarks, are compensated by the ghost fields.
\end{abstract}


\maketitle

\section{Introduction:  Essential Ingredients of  quarkyonic Matter}

The concept of quarkyonic matter~\cite{McLerran:2007qj} was introduced to explain a remarkable feature of QCD in the limit of a large number of colors, $N_c$~\cite{tHooft:1973alw, Witten:1979kh}.  In this limit, fermion loops are suppressed by a factor of $1/N_c$.  This means that deconfinement at finite density cannot occur until a quark chemical potential reaches a value $\mu_Q \sim \sqrt{N_c}  \Lambda_{QCD}$.  This scale can be parametrically large compared to the QCD scale.  For degrees of freedom deep within a Fermi sea, interactions are controlled by exchange interactions which are at a hard momentum scale, and the effect of these interactions should be phenomenologically accounted for by an effectively de-confined quark Fermi sea.  At the Fermi surface, interactions at small angles and small momentum transfers are allowed, because there are states not Fermi blocked above the Fermi surface, and one can scatter into these states.  The effects of confinement are therefore important near the Fermi surface, and the particle degrees of freedom in this region should be thought of as confined nucleons and mesons.

At finite temperature, $T \le \Lambda_{QCD}$, for baryon chemical potentials less than the quark mass $M$, nucleonic degrees of freedom are exponentially suppressed in large $N_c$ by a factor of $e^{-\beta(M-\mu_B)}$.  This means that there are three distinct regions of strongly interacting matter at finite temperature and density.  There is a low temperature and density phase with no baryons that is confined.  There is a high density and low temperature phase that is quarkyonic.  There is a phase at high temperature, or low temperature an ultra-high density that is de-confined.

The picture of zero temperature high density quarkyonic matter is thought of as a Fermi  sea of quarks surrounded by a Fermi shell of nucleons.  There are now models constructed that have this feature arising dynamically~\cite{McLerran:2018hbz, Jeong:2019lhv}.  At low densities it is energetically favorable to have nucleons.  To see this is very simple. The pressure of a non-relativistic nucleon gas is
\begin{equation}
  P^N= \kappa N_{\text{d.o.f.}}^N\frac{(k_F^N)^5}{M_N}
\end{equation}
where $\kappa$ is a numerical factor, which is equal to 4 for an isospin symmetric system,
and  the nucleon Fermi momentum is defined as $k_F^N = \sqrt{\mu_N^2 - M_N^2}$, where $\mu_N$ is the baryon number chemical potential.
For a non-relativistic gas of quarks,
\begin{equation}
  P^Q =\kappa N_c  N_{\text{d.o.f.}}^Q\frac{(k_F^Q)^5}{M_Q}
\end{equation}
where $k_F^Q = \sqrt{\mu_Q^2 - M_Q^2}$. For the quarks, the  chemical potential is $\mu_Q = \mu_N/N_c $, and in the additive quark-nucleon model, the constituent quark mass is $m_Q =  m_N/N_c$. This means that $k_F^N = N_c k_F^Q$, and that  ratio of pressure of quarks to that of baryons is
\begin{equation}
\frac{P^N}{P^Q}  = N_c^3.
\end{equation} 
Therefore when the system is a dilute gas of quarks, and the interaction energy is weak, then the system is entirely in nucleons. At nuclear matter densities $k_F^N \sim \Lambda_{QCD}$,  so that the associated quark Fermi momenta are quite small. One can also verify that the preferred phase at low density is nucleonic by computing the energy per baryon for the same baryon number density of quarks and of nucleons.  One finds that the energy per baryons of a free gas of constituent quarks always exceed that of nucleons.

As the density increases, hard-core nucleon interactions may make the nucleon phase disfavored.  This is because hard-core nucleon interactions are of strength $N_c$, and the Fermi energy of a nucleon can shift by of order of the nucleon mass.   This would naively need to be associated with a huge shift in the density of the baryons, $\rho^N\sim (k_F^N)^3 \sim M_N^3 \sim N_c^3 \Lambda_{QCD}^3$.  This huge increase in density is not realized in quarkyonic matter because the nucleons sit on a shell of decreasing thickness as the Fermi energy of the nucleons increase.  The density of the nucleons should be expected to saturate at the density of matter corresponding to the hard-cores of nucleons, which is of order $\rho_{\textrm{hard core}}\equiv n_0 \sim \Lambda^3_{QCD}$.   Nucleonic matter sits on a shell and generates a density of nuclear matter that is approaching the hard-core density.  As this density is approached, the baryon number density increases by increasing the quark density.  Until the quark Fermi momentum is of order  $k_F^Q \sim \Lambda_{QCD}$, this increase associated with the quarks is quite small.  As the baryon density associated with quarks slowly increases the quark and nucleon Fermi energies rapidly rise until there is a quark Fermi sea with a Fermi energy of order $E^F_Q \sim \Lambda_{QCD}$ and the nucleons become relativistic in the Fermi shell with $E^F_N \sim N_c \Lambda_{QCD} \sim M_N$. Different from the models that exhibit a first order phase transition, quarkyonic matter generates a soft equation of state for small densities, while the rapid increase in the Fermi energies at a slowly varying density leads to a hard equation of state with sound velocities of order one at moderate densities. This is particularly suitable for the neutron star phenomenology~\cite{McLerran:2018hbz, Jeong:2019lhv, Fukushima:2015bda}.

The central problem to deal with in constructing a field theoretical description of quarkyonic matter is that one needs to have both nucleonic and quark degrees of freedom.  An effective field theoretical description of nucleons can only describe matter near the Fermi surface.  The nucleonic degrees of freedom should not be important inside the Fermi sea of quarks as a consequence of  the Pauli blocking. Inside a Fermi sea, the low momentum states are occupied by quarks.  A nucleon is composed of quarks, and therefore cannot propagate when its momentum is $k^N < N_c k^Q$. 

It is useful to have a field theoretical description of quarkyonic matter that is also valid at finite temperature~\cite{Sen:2020peq}.  This involves including pion degrees of freedom, and at least phenomenoligcaly the effect of meson nucleon interactions.  Such a theory should reduce to a theory of nucleons and mesons at low densities, and evolve to quarks at high density and temperature.  It should allow for the possibility of quarkyonic matter.

 It is the purpose of this paper to outline a possible solution to this problem.  The central issue that needs to be resolved is the duality between nucleonic and quark descriptions.  The nucleon can be thought of as a nucleonic state or as an ensemble of quarks.  This means that if quarks occupy low momentum states, the quarks composing a nucleon cannot occupy these same states. Therefore, if we have a field theoretical model, we can have a field that corresponds to a nucleon, and a field that corresponds to quarks, so long is it is constrained so that the quark fields associated with these states do not overlap the same states as are occupied by the quarks.  This can be accomplished by an unconstrained nucleon field, an unconstrained quark field, and a negative metric nucleon ghost field that fill precisely the same state as the quarks, and  whose only purpose is to cancel away the degrees of freedom of the unconstrained nucleon field whose quark states occupy states already occupied by quarks.  

In the paper below, we first argue how such a theory is constructed for free quarks and nucleons.  We then argue how such a theory might be generalized to include interactions with meson fields or in a model where nucleon interactions may be phenomenologically accounted for by an excluded volume. Such a theory may provide a model where one can simultaneously study the onset of quarkyonic matter and the restoration of chiral symmetry.

\section{Ghost and Removing Un-physical States}

	If there is a  Fermi sea of quarks with a chemical potential, $\mu_Q,$ then the nucleon cannot overlap a color singlet state with the quantum number of the nucleon which is composed of quarks.  Such states can exist in the quark Fermi  sea up to a chemical potential $\mu_G \sim N_c \mu_Q$.  The density  of such states is,
\begin{equation}
  \rho^G  = \frac{1}{1 +e^{\beta(N_CE_Q  - \mu_G)}}\,.
\end{equation}
In the additive quark model, $E_N = N_c E_Q$, and if we also think about this color singlet state embedded in the quark Fermi sea as a nucleon, then the energy of this $N_c$ quark state should be thought of as a nucleon energy.  Therefore,
\begin{equation}
 \rho^G  = \frac{1}{1 +e^{\beta(E_N   - \mu_G)}}\,.
\end{equation}

For a noninteracting gas of quarks and nucleons, the density of quarks is
\begin{equation}
  \rho^Q  = \frac{1}{1 +e^{\beta(E_Q -  \mu_Q)}}\,,
\end{equation}
and the density of nucleons, constrained not to propagate in the quark Fermi sea is
\begin{equation}
\rho^N_{\text{const.}}  =\rho^n  =\frac{1}{1 +e^{\beta(E_N - \mu_N)}} - \frac{1}{1 +e^{\beta(E_N   - \mu_G)}}
\end{equation}
This equation is almost trivial since it places the nucleon in a shell of momenta above that of the quark sea, where they will not be Pauli blocked.

We now wish to generalize this description to a field theoretical model.  We have an unconstrained nucleon field with chemical potential $\mu_N$ and mass $M_N$, a ghost nucleon field with chemical potential $\mu_G \sim N_c \mu_Q$ and mass $M_N$ and quark field with mass $m_Q = M_N/N_c$ and chemical potential  $\mu_Q$.  The nucleon field will be denoted by $N$, the quark field will be $Q$ and the ghost field will be $G$. The ghost field will have the same Lorentz structure at the nucleonic field.  It will satisfy anti-periodic boundary conditions in imaginary time, like the nucleons.   It will however be a commuting and not an anti-commuting field.  In a path integral, it will be represented by a c-number integration variable rather than a Grassman algebra variable.  The action for such a theory in Euclidean time is
\begin{eqnarray}
  S_E &  = & \int_0^{\beta} dt \int_V d^3x \left\{
  \overline N \left(\frac{1}{i}\gamma \cdot \partial -i \mu_N \gamma^0   + M_N\right) N  \right. \nonumber \\
 &  & + \overline G \left(\frac{1}{i}\gamma \cdot \partial -i \mu_G \gamma^0   + M_N\right) G \nonumber \\
 &  & \left. +\overline Q \left(\frac{1}{i}\gamma \cdot \partial -i \gamma^0 \mu_Q  + M_Q\right) Q  \right\}\,.\label{action}
\end{eqnarray}

It is useful to define the propagator
\begin{equation}
S(\mu_N, M) = \frac{1}{\frac{1}{i} \gamma \cdot \partial - i \mu_N \gamma^0 +M_N}\,.
\end{equation}
Now if we integrate over a Grassman variable, the path integral for the partition function will give
\begin{equation}
   Z_N = \text{det}^{-1} S(\mu_N, M_N)\,,
\end{equation}
where  the integration over the ghost c-number field yields,
\begin{equation}
   Z_G = \text{det} ~S(\mu_G, M_N)\,.
\end{equation} 
The formula for the grand potential is obtained from
\begin{eqnarray}
\Omega & = & g \frac{1}{\beta V} \text{Tr} ~\Bigl\{\ln(S(\mu_N, M_N)) + N_c\ln(S(\mu_Q, M_Q))\nonumber\\
&&- \ln(S(\mu_G, M_N) )\Bigl\}
\end{eqnarray}
The factor of $N_c$ for the quarks comes from the fact that there are $N_c$ quark fields.  If our quarks and nucleons are isodoublets, the degeneracy factor is $g = 2$.
As expected, the ghost contribution to the action has the opposite sign from that of the nucleons, and is  present to precisely to cancel out the contribution of modes of the nucleon where the quark states of the nucleon are already occupied by quark states.  Since we have chosen the boundary condition in Euclidean time to be precisely the same between the ghost fields and the quark and nucleon fields, except for mass and chemical potential, the determinant is the same.  It is straightforward to evaluate
these determinants by standard methods of diagonalizing in momentum space, and performing a contour integral representation for the Matsubara frequency sum.  The grand potential is obtained as
\begin{eqnarray}
       \Omega &=& -g T \int \frac{d^{3}p}{(2\pi)^3} \left\{ \ln\left[1 + e^{-\beta (E_N(\vec{p})-\mu)} \right] \right.\nonumber\\
       && - \ln\left[1 + e^{-\beta (E_N(\vec{p})- \mu_G)} \right] \nonumber\\
       && + \left. N_c \ln\left[1 + e^{-\beta (E_Q(\vec{p})-\mu_Q)} \right] \right\},\label{gpot}
\end{eqnarray}
where one can notice that the ghosts are present to subtract the contribution of the pressure of the nucleons due to Pauli blocking.
Since the entropy and number density follow by the ordinary thermodynamic relations term by term in the expression above, except for an overall minus sign for the ghost contribution,  all of the expression for the pressure, energy density, entropy and number density are simply the nucleon and quark contributions minus that of the ghost nucleons.

When interactions are included, one can explore various theories to see if one can obtain a reasonable shell structure for quarkyonic matter. As one can find in Eq.~\eqref{gpot} the presssure is obtained in terms of the chemical potentials $\mu_i$, which determine the quasi-particle number densities $n_i$.  Since this quarkyonic matter concept is the theory about the quasi-free quarks and confined quarks (baryon) at the given total baryon number density, the pressure itself is not always the suitable quantity for determination of the matter configurations. For example, if presumed chemical potentials are given, the number densites of the particles (thickness of the shell structure) and subsequent bulk properties can be obtained by extremization of the pressure. On the other hand, if the total baryon number density is given and the chemical potential depends on the particle density ($\mu_i(n_i)$) via the interaction Lagrangian, the shell-like configuration can be determined by extremization of the free energy density ($F/V\equiv f=\epsilon-Ts = -p +\mu n $ where $\epsilon$ and $s$ denotes the energy and entropy density, respectively).

\section{Various Models}

The introduction of ghost fields and a field theoretical description immediately opens up the possibility to explore many different types of theoretical models of quark and nucleon interactions.

\subsection{How to Construct an Excluded-volume Model at $T\rightarrow 0$ limit}

The excluded-volume model~\cite{Jeong:2019lhv} can be derived from the action~\eqref{action}.  First, one can check the shell-like distribution of nucleon in the ideal gas limit by assuming $\mu_G=N_c \mu_Q$. The grand potential~\eqref{gpot} of the symmetric two-flavor system can be written as follows:
\begin{align}
\Omega =-p&= \epsilon_n - \mu_n n_n -(\mu_n- \mu_G) n_G + \epsilon_Q  -\mu_{\tilde{Q}} n_{\tilde{Q}},
\end{align}
where $\tilde{Q}$ represents the quantity counted in a baryon number unit and the relations $\mu_G = \mu_{\tilde{Q}}=N_c \mu_Q$ ($n_G=N_c^3 n_{\tilde{Q}}$), $\mu_n=\mu_N$, and $n_n=n_N-n_G$ are understood. The energy densities are obtained as
\begin{align}
\epsilon_n&=\epsilon_N-\epsilon_G\nonumber\\  
 &=\frac{2}{\pi^2} \int^{(\mu_N^2-m_N^2)^{\frac{1}{2}}}_{(\mu_G^2-m_N^2)^{\frac{1}{2}}}dp p^2 (\vec{p}^{~2}+M_N^2)^{\frac{1}{2}},\\
	\epsilon_Q&=\frac{2 N_c}{\pi^2}  \int^{(\mu_Q^2-m_Q^2)^{\frac{1}{2}}}_{0} dp p^2 (\vec{p}^{~2}+M_Q^2)^{\frac{1}{2}}.
\end{align}
To include the excluded-volume effect ($n_0\neq \infty$), one should apply the enhanced chemical potential of the nucleon ($\mu_N^{\text{ex}}$) and ghost ($\mu_G^{\text{ex}}$) which satisfy following relations:
\begin{align}
n_n^{\text{ex}}&= \frac{n_n}{1-\frac{n_n}{n_0}} = \frac{n_N-n_G}{1-\frac{n_N-n_G}{n_0}}=\frac{2}{\pi^2} \int^{\sqrt{(\mu_N^{\text{ex}})^2-m_N^2}}_{\sqrt{(\mu_G^{\text{ex}})^2-m_N^2}} dp p^2,\\
n_N^{\text{ex}}&=\frac{n_N}{1-\frac{n_n}{n_0}}=\frac{2}{\pi^2} \int^{\sqrt{(\mu_N^{\text{ex}})^2-m_N^2}}_0 dp p^2,\\
n_G^{\text{ex}}&=\frac{n_G}{1-\frac{n_n}{n_0}}=\frac{2}{\pi^2} \int^{\sqrt{(\mu_G^{\text{ex}})^2-m_N^2}}_0dp p^2.
\end{align}
Then, the pressure can be written in terms of the quantities obatined in the reduced system volume:
\begin{align}
p&=-\epsilon_n+ \mu_n n_n +(\mu_n- \mu_G) n_G  - \epsilon_Q +\mu_{\tilde{Q}} n_{\tilde{Q}}\nonumber\\
&=-\epsilon^{\text{ex}}_n   + \mu^{\text{ex}}_n n^{\text{ex}}_n +(\mu^{\text{ex}}_n- \mu^{\text{ex}}_G) n^{\text{ex}}_G  - \epsilon_Q + \mu_{\tilde{Q}} n_{\tilde{Q}},
\end{align}
which is equivalent to the Fermi-Dirac statistics of van der Waals gas without attractive contribution~\cite{Rischke:1991ke, Vovchenko:2015vxa}. Here the modifed version of the constraint $\mu_G=N_c \mu_Q$ is applied:
\begin{align}
{k_F^G}^{\text{ex}}=\left((\mu_G^{\text{ex}})^2-m_N^2\right)^{\frac{1}{2}} = N_c k_F^Q = N_c(\mu_Q^2-m_Q^2)^{\frac{1}{2}}\,,
\end{align}
which leads to $n_G^{\text{ex}} =N_c^3 n_{\tilde{Q}}$. The energy densities are given by
\begin{align}
\epsilon^{\text{ex}}_n&=\frac{2}{\pi^2} \int^{\sqrt{(\mu_N^{\text{ex}})^2-m_N^2}}_{\sqrt{(\mu_G^{\text{ex}})^2-m_N^2}}dp p^2 (\vec{p}^{~2}+M_N^2)^{\frac{1}{2}},\\
\epsilon_Q&=\frac{2N_c}{\pi^2}  \int^{\sqrt{\mu_Q^2-m_Q^2}}_{0} dp p^2 (\vec{p}^{~2}+M_Q^2)^{\frac{1}{2}}.
\end{align}
From the trivial relations
\begin{align}
\frac{\partial \epsilon^{\text{ex}}} {\partial n^{\text{ex}}}&=\mu^{ex } =\left( 1 - \frac{n}{n_0} \right) \mu + \frac{\epsilon}{n_0},\\
\epsilon^{\text{ex}}& =\frac{\epsilon}{ 1 - \frac{n}{n_0}},
\end{align}
all the quantities for the total system (denoted without the superscript `ex') can be obtained from the potential~\eqref{gpot} expressed in terms of the quantities calculated in the reduced system volume (denoted by the superscript `ex') :
\begin{align}
\mu_n & = \frac{\partial \epsilon}{\partial n_n} =\frac{\mu^{\text{ex}}_n}{\left( 1 - \frac{n_n}{n_0} \right)}  - \frac{\epsilon_n^{\text{ex}}}{n_0},\\
\mu_{Q} & =\frac{\partial \epsilon}{\partial n_{\tilde{Q}}}= \mu_{\tilde{Q}} + N_c^3\left( 1 - \frac{n_n}{n_0} \right)(\mu^{\text{ex}}_n- \mu_{\tilde{Q}}),\label{qc}
\end{align}
where $\epsilon=\epsilon_n+ \epsilon_Q $. In practical calculation, the quark chemical potential~\eqref{qc} needs additional cut-off factor $N_c^3\rightarrow N_c^3 (\frac{3\pi^2}{2}n_{\tilde{Q}})^{\frac{1}{3}}/\sqrt{(\frac{3\pi^2}{2}n_{\tilde{Q}})^{\frac{2}{3}}+\Lambda^2}$ to satisfy the physical baryon number conservation ($n_B=n_n+n_{\tilde{Q}}$ with $n_n,n_{\tilde{Q}} \geq 0$)~\cite{Jeong:2019lhv}. Although the action~\eqref{action} leads to the grand potential $\Omega(T\rightarrow0,~V,~\mu)=-pV$, it is easier to work within the free energy $F(T\rightarrow0,~V,~N)=E=-pV+\mu N$ since the equilibrium will be determined by additional constraint which requires the extremum of $F(T\rightarrow0,~V,~N)/V=\epsilon$ at the fixed total baryon number density ($n_B=n_n+n_{\tilde{Q}}$ with fixed $V$):
\begin{align}
d\epsilon &= \frac{\partial \epsilon}{\partial n_{n}} d n_{n} +  \frac{\partial \epsilon}{\partial n_{\tilde{Q}}} d n_{n_{\tilde{Q}}}\nonumber\\
&= \left( \frac{\partial \epsilon}{\partial n_{n}}- \frac{\partial \epsilon}{\partial n_{\tilde{Q}}} \right) d n_{n} =0,
\end{align}
where  $d n_B=dn_n+dn_{\tilde{Q}}=0$ is understood. In this step, the chemical potentials are not free parameters but dependent on $n_n$ and $n_{\tilde{Q}}=n_B-n_n$. According to the shell-like distribution of the nucleons, $\mu_n$ does not decrease monotonically~\cite{Jeong:2019lhv}, which means that $\mu_n n_n+\mu_{\tilde{Q}} n_{\tilde{Q}}=\mu_B n_B$ (where $\mu_n=\mu_{\tilde{Q}}=\mu_B$)  does not correspond to the extremum of $\mu_n n_n+\mu_{\tilde{Q}} n_{\tilde{Q}}$ exactly whenever $n_{\tilde{Q}}\neq0$.
Therefore, once the low phase space is saturated by the quasi-free quark sea ($n_{\tilde{Q}}\neq0$), the quarkyonic-like configuration determined at the minimum of the free energy does not always correspond to  the extremum of pressure in variations of $n_{\tilde{Q}}$.

\subsection{How to Construct a Quarkyonic Sigma Model}

Pions may be introduced into this theory in a controlled way so long as the density and Fermi momentum scales of the nucleons are not too large.  We do this by replacing the masses above by the couplings to a scalar sigma field and pion field as
\begin{eqnarray}
   M_N&  = & g_N( \phi + i \pi \gamma^5) \nonumber \\
   M_Q&  = & g_Q( \phi + i \pi \gamma^5)
\end{eqnarray}
In the additive quark-nucleon model $g_N = N_c g_Q$.
One could also implement variants of the theory above as a Nambu--Jona-Lasinio model by introducing polynomials of the nucleon and ghost field.  There are many possibilities.  Notice that in this construction, ghost and nucleon couplings to the pion and sigma fields are identical.

\subsection{The Nucleonic Interactions}

One can introduce various interactions for nucleons through meson fields.  Because the ghosts are present to precisely cancel the effect of nucleons in the Fermi sea, coupling to these meson fields are identical, and they couple to the $\overline N \Gamma_i N + \overline G_i\Gamma G$, where $\Gamma_i$ represent the combination of Lorentz Gamma and flavor matrices  coupling to the meson field denoted by $i$.  If the meson interaction is replaced by a contact interaction, then the coupling to fermions is through the sum of nucleon and the sum of ghosts.  One can imagine many variants of such a constructions where there are effective meson interactions that involve coupling to quarks.

\subsection{Quark Interactions}

There are various ways to implement quark interactions.  Their interactions among themselves might be described in QCD by introducing gluons, or in an effective theory involving Wilson line interactions~\cite{Pisarski:2000eq}.

\section{Summary and Conclusions}

The quarkyonic model above allows for a dynamical study of the formation of a Fermi shell of nucleons.  One simply computes the pressure, requires that the total baryon number is fixed, and then searches for an extremum as a function of the quark chemical potential.  It is known that in an excluded volume model,  such an extremum exists, and there is good reason to expect this in generic field theory models of nucleon interactions.  Nevertheless, this needs to be carefully investigated.

There are many theoretical questions which may now be addressed.  How does chiral symmetry restoration affect the formation of quarkyonic matter?  How does one include the effects of nucleon interactions to make a viable model of nuclear matter that properly matches on to known physics at nuclear matter density and below?  How does quarkyonic matter appear in the $T, \mu_B$ plane? How does confinement appear within this phase diagram?    What form of nucleonic interactions can generate a phenomenologically successful theory of quarkyonic matter?

\begin{acknowledgments}
Dyana C. Duarte, Saul Hernandez-Ortiz, Kie Sang Jeong, and L.  McLerran was supported by  the U.S. DOE under Grant No. DE-FG02- 00ER4113.  Dyana C. Duarte, Saul Hernandez-Ortiz, Kie Sang Jeong acknowledge the support from the Simons Foundation on the Multifarious minds Grant No. 557037 to the Institute for Nuclear Theory.  Larry McLerran acknowledges useful insight provided by Jin-feng Liao, Rob Pisarski and Sanjay Reddy.

\end{acknowledgments}

\end{document}